\documentclass{emulateapj}
\shorttitle{Differential element abundances in NGC~6397}
\shortauthors{Koch \& McWilliam}
\begin{document}
\title{A differential chemical element analysis of the metal poor Globular Cluster NGC~6397\altaffilmark{$\dagger$}}
\author{Andreas Koch\altaffilmark{1} \& Andrew McWilliam\altaffilmark{2}}
\altaffiltext{$\dagger$}{This paper includes data gathered with the 6.5 m Magellan Telescopes located at Las Campanas Observatory, Chile.}
\altaffiltext{1}{Zentrum f\"ur Astronomie der Universit\"at Heidelberg, Landessternwarte, Heidelberg, Germany}
\altaffiltext{2}{Observatories of the Carnegie Institution of Washington, Pasadena, CA, USA}
\email{akoch@lsw.uni-heidelberg.de, andy@obs.carnegiescience.edu}
\slugcomment{Accepted for publication in the Astronomical Journal}

\begin{abstract}
We present chemical abundances in three red giants and two turn-off stars in the metal poor Galactic globular cluster (GC) NGC~6397 
based on   spectroscopy obtained with the MIKE high resolution spectrograph on the Magellan 
6.5-m Clay telescope.  Our results are based on a line-by-line differential abundance analysis
relative to the well-studied  red giant Arcturus and the Galactic halo field star Hip~66815. 
At a mean of $-2.10\pm0.02$ (stat.) $\pm0.07$ (sys.)  
the differential iron abundance is in good agreement with other studies in the literature 
based on $gf$-values. As in previous, differential works we find a distinct departure from 
ionization equilibrium in that the abundances of \ion{Fe}{1} and \ion{Fe}{2} differ by 
$\sim$0.1 dex, with opposite sign for the RGB and TO stars.  
The $\alpha$-element ratios are enhanced to 0.4 (RGB) and 0.3 dex (TO), respectively, and we also confirm
strong variations in the O, Na, and Al/Fe abundance ratios. Accordingly, the light-element abundance patterns in one of the red giants can be attributed to   
pollution by an early generation of massive SNe~II. 
TO and RGB abundances are not significantly different, with the possible exception of Mg and Ti, which is, however, amplified by the patterns in one TO 
star, additionally belonging to this early generation of GC stars. 
We discuss interrelations of these light elements as a function of the GC metallicity. 
\end{abstract}
\section{Introduction}
Globular clusters (GCs) are amongst the oldest stellar aggregates in the Universe and therefore  bear the traces of the earliest phases during which the  
Galaxy was assembled. 
While the Milky Way's GC system appears homogeneous (e.g., Koch \& C\^ot\'e 2010, and references therein) and similar to halo field stars 
in many regards  (Geisler et al. 2007), a number of characteristics are clearly at odds with the stellar halo (such as variations of the light chemical elements 
O, Na, Al; Gratton et al. 2004). Moreover, the second-parameter effect of remote GCs (i.e., variations of their horizontal branch morphology at a given metallicity) and  
the dual nature of stellar halos (Koch et al. 2008; Carrolo et al. 2010) emphasize the need to couple an accurate age- and distance scale for Population~II objects with their chemical properties. 
This requires, however, an accurate, absolute abundance scale as well and, e.g., an age measurement to within 5\% accuracy is only achievable once the clusters' metallicities (accounting for Fe and $\alpha$-element abundances) are known to within $\sim$0.05 dex. 

In the first two papers of a series  (Koch \& McWilliam 2008; Koch \& McWilliam 2010; hereafter
KM08 and KM10), we have initiated such a new GC scale. With NGC~6397 the present work concludes
the establishment of this scale:  NGC~6397 is a metal-poor system and therefore the three GCs (47~Tuc, M5, NGC~6397),
and the reference star Arcturus, that define our scale cover a broad, representative range of $\sim$1.5 dex in [Fe/H]. 

NGC~6397 is the second closest GC (d$_{\odot}$=2.3 kpc; d$_{\rm GC}$=6.0 kpc; Harris 1996 [2010 edition]) to the observer and thus an optimal target for abundance studies of the inner halo component. 
% d_GC = 6 kpc
%
Previous studies have established NGC~6397 as a metal poor system below [Fe/H]$\la -2$ dex, 
with clear evidence of variations in those light elements that are 
affected by proton-capture nucleosynthesis; in particular, Carretta et al. (2009a,b) found a pronounced Na-O
anti-correlation.  Iron-peak and the $\alpha$-elements were, however, reported
to show the degree of homogeneity that is nowadays a definitive characteristic of Galactic
GCs (Castilho et al. 2000; Carretta et al. 2009c; Lind et al. 2011), with the possible exception
of abundance variations due to atomic diffusion and mixing acting in the hotter stars (Korn et al. 2007; Lind et al. 2008). 

This paper is organized as follows; In \textsection 2 we present the data set and the standard reductions 
taken, while our atomic linelist, stellar atmospheres are briefly introduced in \textsection 3, where we 
also recapitulate our differential abundance analysis relative to  Arcturus and discuss in depth the question of ionization equilibria. 
Our metallicity scale,  abundance errors and results are  
discussed in \textsection 4. A brief note on mass-loss from the GC stars is given in \textsection 5. Finally, \textsection 6 summarizes our findings. 
\section{Data \& Reduction}
Observations were carried out over five nights in May--June 2005  
using the Magellan Inamori Kyocera Echelle (MIKE) 
spectrograph at the 6.5-m Magellan2/Clay Telescope (see the observing log in Table~1). 
\begin{deluxetable*}{rcccccccccc}
\tabletypesize{\scriptsize}
\tablecaption{Details of observations and target properties}
\tablewidth{0pt}
\tablehead{
 \colhead{ID\ }           &     \colhead{Type}   &     \colhead{Date}      &   \colhead{Exp.}  &
  \colhead{S/N}           &  \colhead{$\alpha$}  &  \colhead{$\delta$}     &   \colhead{$V$}   &
 \colhead{$B-V$}          &   \colhead{$V-K$}    &   \colhead{$v_{\rm HC}$}  \\
                          &                      &                         &  \colhead{[s]}    &
 \colhead{[pixel$^{-1}$]} &  \colhead{(J2000.0)} &  \colhead{(J2000.0)}    &  \colhead{[mag]}  &
 \colhead{[mag]}          &   \colhead{[mag]}    &  \colhead{[km\,s$^{-1}$]} 
}
\startdata
13414 & RGB & 2005 Jun\/\, 03  	   & 3600  & \llap{6}20 & 17:40:20.2 & $-$53:42:01.8 &  \phantom{1}9.89 & 1.47 &  3.61 & 20.7     \\
8958  & RGB & 2005 May 31  	   & 4230  & \llap{5}80 & 17:40:38.9 & $-$53:45:25.2 & 10.28 & 1.32 &  3.32 & 24.4    \\
7230  & RGB & 2004 May 04  	   & 9900  & \llap{3}80 & 17:40:43.2 & $-$53:44:45.0 & 10.34 & 1.35 &  3.35 & 20.0      \\
13552 &  TO & 2005 May 31  	   & \llap{1}0800 &  50 & 17:40:19.3 & $-$53:38:07.8 & 16.26 & 0.58 & 1.77 & 15.5  \\
365   &  TO & 2005 Jun\/\, 02        & 7200  &  40 & 17:40:19.4 & $-$53:43:08.3 & 16.27 & 0.58 & 1.77 & 18.9 
\enddata
\end{deluxetable*}
The targets for this project were selected from the catalogue of Kaluzny (1997), from which 
we adopt the identification numbers. Infrared colors were taken from the Two Micron All Sky Survey 
(2MASS; Skrutskie et al. 2006). 
In practice, we chose red giants that have photometric temperatures within $\pm$100 K of that of Arcturus, 
as well as turnoff (TO) stars similar to the well studied halo field reference star Hip 66815 (KM08), 
which will facilitate our differential analysis. 
Fig.~1 shows the location of our targets on a color magnitude diagram (CMD) using the Kaluzny (1997) data\footnote{We note, however, that Kaluzny (1997) cautions that his data  ``were not aimed of getting a deep and accurate photometry suitable for study of the cluster properties.''.}. 
A list of the targeted stars is given in Table~1, together with their  photometric properties. 
\begin{figure}[!ht]
\begin{center}
\includegraphics[angle=0,width=0.9\hsize]{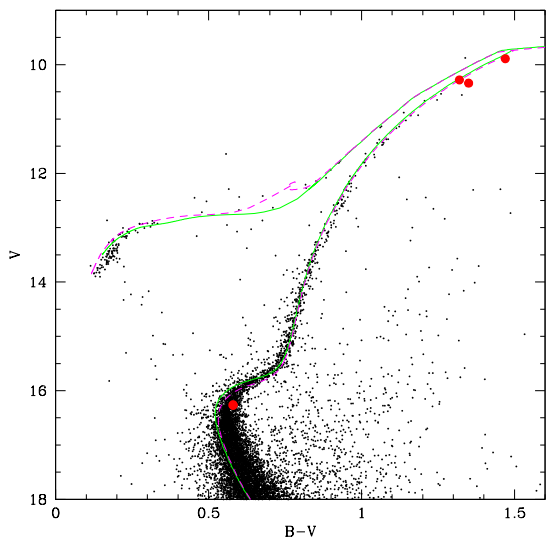}
\end{center}
\caption{
Color-magnitude diagram of our targets (red symbols) from the photometry of Kaluzny (1997).  
 Also indicated are Teramo isochrones for 12 Gyr (solid green lines) and 13 Gyr (dashed magenta lines),
$\alpha$-enhanced, $\eta=0.4$, with [Fe/H]=$-2.14$, shifted to E(B$-$V)=0.18 and a distance modulus of
12.03 mag. Note that we actually observed two subgiants with similar colors that overlap in this figure.}
\end{figure}

None of our targets coincides with the spectroscopic studies of Castilho et al. (2000), 
Korn et al. (2007), Lind et al. (2008, 2011), nor with Carretta et al. (2009a,b,c). 
All the above studies focused on various evolutionary stages, from TO to RGB, addressing different questions and different chemical elements (chiefly the Li-abundances and the Na-O anti-correlation) in a total of some 70--80 cluster stars. 
Moreover, all studies performed element analyses using laboratory $gf$ values 
so that it appears timely to place the stars in NGC~6397 on a homogeneous differential scale and to establish a firm comparison of the GC's mean abundance patterns from both the differential and the absolute approach. 

For the present data, we utilized both the red and blue echelle 
set ups, which yield a wavelength range of 3500--9300\AA. 
By using a slit with a width of 0.5$\arcsec$ and a CCD binning of 2$\times$1 pixels, 
we obtained a spectral resolution of $R\sim$40000. 
Typically, each star was observed for 45--60 minutes, 
where we split the observations into several exposures to facilitate cosmic ray removal. 
The seeing typically ranged from 0.5--0.9$\arcsec$.

The data reduction proceeded in exact analogy to that described in KM08, i.e., by using the 
pipeline reduction software by Kelson et al. (2003) and subsequent continuum normalization using 
a  high-order polynomial fit.  
Typically, our observing and reduction strategy results in S/N ratios of  several hundreds per extracted
pixel for the red giants (Table~1), 
as measured in the order containing the H$\alpha$-line and still as high as 200 per pixel at 5000\AA. For
the TO stars, the S/N ratios reach 50 per pixel on average and we chose to chiefly work on the blue
part of the data (Section~3.1).

These high S/N ratios enabled us to measure accurate equivalent widths (EWs) down to the 1 m\AA~level, which
ensures that we can measure the same transitions as in our previous studies, despite the much lower
metallicity of NGC~6397 compared to the previous targets. 

Radial velocities of the target stars were determined from the Doppler shifts of typically 35  
strong, unsaturated and unblended absorption features. Variations of the individual 
velocities are, however, not critical for our analysis, since the center of each spectral line 
is independently determined during our later EW measurement process. 
Overall, we find a mean heliocentric radial velocity of 
19.9$\pm$1.2 km\,s$^{-1}$\ with a
dispersion of 2.8$\pm$0.9 km\,s$^{-1}$  		% ALL
(see Table~1), which is slightly higher than the value of $\sim18.8$ found in the literature, yet agrees to within the uncertainties  (Meylan \& Mayor 1991; Milone et al. 2006; Lind et al. 2009).  
\section{Abundance analysis}
For the present analysis we proceed in exact analogy to the methods outlined in detail 
in KM08 and KM10. Here, we briefly summarize our differential abundance approach.
\subsection{Line list}
In order to maintain our goal of a homogeneous abundance scale we opted to use the 
identical line list as in the previous works. The line data were taken from Fulbright et al. 
(2006) for the iron lines and Fulbright et al. (2007) for the $\alpha$-elements. 
These authors had assembled their list such to exclude any contamination from blended features. 
At an estimated mean [Fe/H] of below $-$2 dex (e.g., Carretta et al. 2009),  NGC~6397 is more
metal poor than M 5 by approximately 0.7 dex (KM10) and metal deficient by a factor of 20 
compared to the metal rich 47 Tuc (KM08).  Thus most of the transitions used for our measurements
in those GCs will be very weak.  Moreover, our chosen reference star, Arcturus, has a 
significantly higher iron abundance ([Fe/H]=$-0.5$ dex; Fulbright et al. 2006).  However, the
present, high S/N red giant spectra allow us to still measure the majority of lines, at a few
m\AA, to a high accuracy.  

The reason for choosing relatively weak lines in Arcturus that are very weak in the program GC stars 
is that these lines are unsaturated, and so have the maximum sensitivity to abundance.  In addition,
systematic errors in the adopted damping constants and microturbulent velocity parameter do not
affect the abundance result for such weak lines.  In this way, these lines offer excellent advantage
for measuring precise abundances.
In fact, approximately 80 \ion{Fe}{1} lines of sufficient strength could be 
reliably measured in the giants, allowing for an accurate abundance determination. 
Also the  absorption lines of the heavier elements from the line lists of KM08 and KM10 
are clearly visible in the spectra so that we did not attempt to supplement our line lists from other sources. 

In practice, EWs of the absorption lines were derived using the semi-automated code GETJOB 
(McWilliam et al. 1995), based on a Gaussian fit to the line profiles.
In this way we could achieve 
typical EW uncertainties of  $\la$5\%, as determined from the r.m.s. scatter around 
the best-fit profile. 
The same lines on adjacent spectral orders 
agree to within this uncertainty and as the final value we adopted the error-weighted mean. Line-free continuum regions were adopted from the study of 
Fulbright et al. (2006), based on a detailed study of the metal rich giant $\mu$Leo.  

On the other hand, none of the absorption lines in the red spectra range could be used for the much fainter TO stars
and also the line list used for the metal rich 47~Tuc TO stars of KM08 contains no measurable feature in the NGC~6397 stars. 
Therefore we compiled a new TO line list based on measurements of unblended transitions in regions with well defined continua in our reference 
dwarf star, Hip~66815, chiefly making use of the blue part of the echelle from $\sim$3500--5000\AA. 
This way, we could identify $\sim$50 useable \ion{Fe}{1} lines in the metal poor TO stars. 

Neither hyperfine structure splitting for Na nor Al was included in our analyses.
The final lists with our measured EWs are shown in Tables~2 and 3. 
\begin{deluxetable}{cccccc}%[!hbt]
\tabletypesize{\scriptsize}
\tablecaption{Line list for the red giants}
\tablewidth{0pt}
\tablehead{ \colhead{} & \colhead{$\lambda$} &  \colhead{E.P.} &  \multicolumn{3}{c}{Equivalent Width [m\AA]} \\
\cline{4-6}
 \raisebox{1.5ex}[-1.5ex]{Ion} & \colhead{[\AA]} &  \colhead{[eV]} & \colhead{13414} & \colhead{8958} & \colhead{7230}  }
\startdata
\ion{Fe}{1}  &  5432.95  &  4.45  &   28.9  &   22.9  &   26.8       \\
\ion{Fe}{1}  &  5460.87  &  3.07  &   3.2   &   1.4   &   2.9	     \\
\ion{Fe}{1}  &  5462.96  &  4.47  &   44.0  &   36.4  &   38.4	     \\
\ion{Fe}{1}  &  5466.99  &  3.57  &   9.7   &   7.7   &   8.0	     \\
\ion{Fe}{1}  &  5470.09  &  4.45  &   4.0   &   2.6   &   4.8	     
\enddata
\end{deluxetable}
\begin{deluxetable}{ccccc}%[!hbt]
\tabletypesize{\scriptsize}
\tablecaption{Line list for the turn-off stars}
\tablewidth{0pt}
\tablehead{ \colhead{} & \colhead{$\lambda$} &  \colhead{E.P.} &  \multicolumn{2}{c}{Equivalent Width [m\AA]} \\
\cline{4-5}
 \raisebox{1.5ex}[-1.5ex]{Ion} & \colhead{[\AA]} &  \colhead{[eV]} & \colhead{13552}  & \colhead{365}  }
\startdata
 \ion{Fe}{1} &    3552.83 &  2.88 &    20.1 &  \nodata   \\
 \ion{Fe}{1} &    3640.39 &  2.73 &  \nodata &    15.8   \\
 \ion{Fe}{1} &    3760.05 &  2.40 &    15.2 &  \nodata   \\
 \ion{Fe}{1} &    3786.68 &  1.01 &    24.6 &  \nodata   \\
 \ion{Fe}{1} &    3790.09 &  0.99 &    35.6 &  \nodata   
\enddata
\tablecomments{Tables 2 and 3 are published in entirety in the electronic edition of the {\it Astronomical 
Journal}. A portion is shown here for guidance regarding their form and content.}
\end{deluxetable}
\subsection{Stellar atmospheres and parameters}
Stellar abundances for each of the absorption lines were computed using the {\em abfind} 
driver of the 2010 version of the synthesis program MOOG (Sneden 1973). 
The stellar atmospheres for this analysis were generated from the 
Kurucz LTE models\footnote{\url{http://kurucz.harvard.edu}} without convective overshoot. Since both Arcturus and NGC~6397  are enhanced in the $\alpha$-elements by approximately $+0.4$\,dex 
(Arcturus: Fulbright et al. 2007) and $\sim0.3$ dex (NGC~6397: Castilho et al. 2000; Lind et al. 2010), 
 the  $\alpha$-enhanced opacity distributions AODFNEW\footnote{\url{http://wwwuser.oat.ts.astro.it/castelli}} by F. Castelli were used in our analysis.
For Arcturus we resumed the same EWs and abundances for  individual lines as in KM08,  
based on its well-established atmospheric parameters (T$_{\rm eff}$=4290 K, log $g$=1.64, $\xi$=1.54 
km\,s$^{-1}$, [$M$/H]=$-$0.49 dex). 
\subsubsection{The red giant sample}
Photometric effective temperatures for the  red giants in NGC~6397  were calibrated  from the empirical  V$-$K color-temperature  
relation of Alonso et al. (1999). To this end, we combined the V-band photometry of 
Kaluzny (1997) with infrared K-band magnitudes from 2MASS (Skrutskie et al. 2006).  
In order to convert the photometry into the TCS system, needed for the Alonso et al. (1999)-calibrations, 
we applied the transformations from Alonso et al. (1998) and Cutri (2003). 

For the reddening we adopt 0.18$\pm$0.01 mag as fiducial value from many recent studies (Richer et al. 2008, and references therein).    
In addition, we use the extinction law of Winkler (1997) throughout this work.
As a result, the uncertainty on T$_{\rm eff}$ from photometric and the calibration errors, is of the order of 30 K.
As noted in KM08, in a differential analysis it is necessary to add a zero point shift of +38 K 
to the GC giant temperatures, due to a difference in the Alonso et al. (1999) color calibration for 
Arcturus compared to its value from angular diameter measurements.  This is consistent with
recent re-determinations of the (V$-$K)--T$_{\rm eff}$ relations by Gonz\'alez Hernandez \& 
Bonifacio (2009), who found tempertures for the RGB stars higher than the Alonso et al. (1999) scale
by $\sim$54K.

We note, however, that while our zero-point adjustment is accurate for Arcturus, for RGB stars
with [Fe/H] very far from that of Arcturus the correction to the Alonso et al. (1999) 
(V$-$K) temperature-color relation may be different.  

Secondly, we selected the spectroscopic temperature for each star that gave a zero slope 
for the trend of (differential) \ion{Fe}{1} abundance versus excitation potential.  
As each abundance point in NGC~6397 is truly differential to the same absorption line in 
Arcturus, our spectroscopic temperatures are on the same physical T$_{\rm eff}$ scale 
as the reference star. 
On average, the difference between excitation- and photometric temperatures, T(V$-$K)$-$T(spec), 
is  91 K with an 
r.m.s. scatter of 30 K % 30.4
so that the random error on the mean difference from the three RGB stars is 17.5 K. % = sigma(delta T) / sqrt(3)
Since a reddening change of 0.01 in E(B$-$V) induces  
a shift in temperature of $\sim$19 K, a reddening error of 0.05 mag would need to be invoked to account for 
 the T$_{\rm eff}$ differences, which is unlikely given the accurate determination of this parameter from deep CMDs.
In practice, we used the average of both estimates in the subsequent abundance analyses. 

We note that Kaluzny (1997) stated that his photometry for NGC~6397 was not aimed at acquiring deep and accurate
photometry suitable for study of the cluster properties, and this limitation may be the reason why the main-sequence
is not well matched in Fig.~1.
In order to check the accuracy of the Kaluzny (1997) photometry we compared to the studies of Anthony-Twarog, 
Twarog \& Suntzeff (1992), and Alcaino \& Liller (1980).  Stars common to all three studies showed very similar mean 
magnitudes, although there was significantly greater dispersion in the Alcaino \& Liller (1980) data.
The correlation between Anthony-Twarog et al. (1992) and Kaluzny (1997) photometry was excellent, with the
Kaluzny (1997) V values smaller by 0.041$\pm$0.007 mag.  If we shift the Kaluzny (1997) V photometry onto
the  Anthony-Twarog et al. (1992) scale, our photometric T$_{\rm eff}$ values would decrease by 24 K, and the
difference between photometric and spectroscopic T$_{\rm eff}$ would be reduced to 67 K, partially resolving the
discrepancy between photometric and spectroscopic temperatures.  However, this would reduce the adopted mean
temperatures by only 12 K, resulting in a 0.01 dex decrease in our final [Fe~I/H] values.

Surface gravities for the GC stars were derived from the stellar structure equations 
(Eq.~1 in KM08) and assume the  previously determined T$_{\rm eff}$ and  standard Solar values.  
To obtain luminosities we used the dereddened V-band photometry of Kaluzny (1997)
with bolometric corrections from the Kurucz database and a distance modulus to NGC~6397 of 
(m$-$M)$_0$=12.03$\pm$0.10 mag. The latter is an average from the studies of   
Reid \& Gizis (1998); Gratton et al. (2003); Hansen et al. (2007), and Richer et al. (2008)  and based on various methods   
such as fits to the subdwarf or white dwarf sequences. 

Stellar masses of the red giants stars are based on a comparison with the most recent Teramo 
$\alpha$-enhanced isochrones (Pietrinferni et al. 2004) with $\alpha$ element opacities from Ferguson et al. (2005). 
This is illustrated in Fig.~1, where we compare the observed CMD, from Kaluzny (1997), with
12 Gyr and 13 Gyr $\alpha$-enhanced, $\eta=0.4$, Teramo isochrones, at [Fe/H]=$-$2.14  dex.
In Fig.~1 the isochrones pass somewhat to the blue of the bulk of the stars on the main-sequence,
although within the scatter of the points, and the RGB and AGB branches appear well matched;
however, the bluest BHB stars on the horizontal branch (HB) are better matched by the 
13 Gyr isocrhone than the 12 Gyr isochrone.  Unfortunately, we have no way to investigate possible
zero-point corrections for the B-band data of Kaluzny (1997), so we refrain from an attempt to estimate
the age of NGC~6397 from this B,V data.

We adopt an age for NGC~6397 of 12.0 Gyr, based on color-magnitude diagram studies by 
Anthony-Twarog \& Twarog (2000) and Gratton et al. (2003), and the white dwarf cooling results of
Hansen et al. (2007), increased by 0.5 Gyr for the for Ly-$\alpha$ opacity correction suggested by
Kowalski (2007).  This age is in agreement with Kaluzny et al. (2008), who adopted 12.0$\pm$0.5 Gyr.

The 12 Gyr, $\alpha$-enhanced, [Fe/H]=$-$2.14 Teramo isochrone was interpolated, giving stellar masses 
of 0.66 M$_{\sun}$ for the RGB stars and 0.73 M$_{\sun}$ for the TO stars; if our giant stars are, instead, 
on the AGB, the mass is decreased to 0.64 M$_{\sun}$, which would decrease log\,$g$ by only 0.01 dex. 
For more metal-rich isochrones the RGB masses increase, at roughly $+$0.02 M$_{\sun}$ for
$\Delta$[Fe/H]=$+$0.10 dex, while the AGB masses decrease by a mere 0.003 M$_{\sun}$.  For an 11 Gyr isochrone
the masses increase by 0.03 M$_{\sun}$, while for 13 Gyr the masses are decreased by 0.02 M$_{\sun}$.
Thus, we estimate the total uncertainty in log\,$g$ due to the isochrones at $\pm$0.02 dex.
Accordingly the log $g$ values given in 
Table~4  could be determined to within $\pm$0.06 dex, based on individual uncertainties 
in T$_{\rm eff}$, mass, and the main contribution from the uncertainty in the distance modulus of $\sim$0.10 mag.    

Next, the microturbulent velocity $\xi$ was initially set to that of Arcturus (1.54 km\,s$^{-1}$), 
and iterated by eliminating any slope in the plot of differential abundance 
with EW.  A linear fit to the data then fixed $\xi$ to within 0.05 km\,s$^{-1}$. 
Finally, the metallicities of the atmospheres [$M$/H] were equated to the [\ion{Fe}{1}/H] abundance 
from the previous iteration step. Table~4 lists the final stellar parameter set that defined our atmospheres 
for the abundance analysis. 
%
%
%	Note -- T(V-K) in table are  differential to Arcturus and include the 38K offset!
%	
%
\begin{deluxetable}{rccccc}%[!bt]
\tabletypesize{\scriptsize}
\tablecaption{Atmospheric Parameters}
\tablewidth{0pt}
\tablehead{ \colhead{ID} & \multicolumn{3}{c}{T$_{\rm eff} [K]$} & \colhead{log\,$g$}  & \colhead{$\xi$} \\
\cline{2-4}
 & (V$-$K) & (spec.) & (average) & [cm\,s$^{-2}$] & [km\,s$^{-1}$]}
\startdata
13414 &  4167 & 4081 & 4124 & 0.29 & 1.74  \\
 8958 &  4355 & 4231 & 4293 & 0.63 & 1.70  \\
 7230 &  4330 & 4266 & 4298 & 0.66 & 1.56  \\
13552 &  6253 & 6250 & 6250\rlap{$^{\rm a}$} & 3.95 & 1.18  \\
  365 & 6249 & 6250 & 6250\rlap{$^{\rm a}$} & 3.95 &   1.02
\enddata
\tablenotetext{a}{Average including T$_{\rm eff}$ (H$\alpha$) and the other color indices.}
\end{deluxetable}
\subsubsection{The Turnoff sample}
 For the unevolved, fainter TO stars, the K1.5~{\sc III} star Arcturus does not provide a  
realistic reference; rather, it is necessary to perform differential analysis relative to a star
with atmosphere parameters similar to the TO targets.  Hence, we follow KM08 in measuring abundances
on a line by line basis differential to the Galactic halo field dwarf Hip 66815, for which 
we adopt the stellar parameters derived differentially in KM08: T$_{\rm eff}$=5812 K, log $g$=4.41,
$\xi$=1.13 km\,s$^{-1}$, and [Fe {\sc i}/H]=$-0.76$ dex.    Abundances for Hip~66815 were measured
differentially, relative to the sun, and was facilitated by our very high S/N spectrum of Hip~66815,
which permitted measurement of very weak lines that are also unsaturated in the solar spectrum.
 
Photometric temperatures for the TO sample are based on their B$-$V, V$-$H, V$-$J, and V$-$K colors and are derived from the dwarf calibrations of Ram\'irez \& Mel\'endez (2005).   
Owing to the large uncertainties on the 2MASS magnitudes of the order of 0.10 mag for the faint stars, the values from the four  indicators show a broad 1
$\sigma$ scatter of 170 K. While the B$-$V and V$-$K values are in good agreement, the H- and J-based color-temperatures yield much higher values, albeit with large uncertainties. The error-weighted mean T$_{\rm eff}$ (phot.) is thus 6270 K, 6310 K, resp., for the two TO stars. 

Secondly, we obtained an independent estimate of T$_{\rm eff}$ from a fit of synthetic spectra
with varying temperatures to the H$\alpha$ line profiles (Gratton et al. 2005; KM08), which
yields  reddening-free temperatures of 6250$\pm$150 K for either star, in good agreement with
the above photometric values.  In order to account for damping effects such as Stark and van
der Waals broadening in the dense atmospheres of the dwarfs, we multiplied the damping constants
in the classical Uns\"old approximation by a constant factor of 6.5, both for H$\alpha$ and for
all subsequent metal absorption line studies.  

Finally, as for the red giants above we also estimate the spectroscopic temperatures from the excitation plot. For each star, a spectroscopic T$_{\rm eff}$ of 6250$\pm$150 K is able to reproduce a reasonably flat slope of abundance against excitation potential. Since this is in good agreement with the overall values from the methods described above, we adopt an average  T$_{\rm eff}$ 6250$\pm$150 K as the atmospheric temperature of our two TO stars. 

The TO stars' gravities are then based on the same isochrone tracks as above, which imply a stellar mass of 0.79 M$_{\odot}$. 
As before, the uncertainties on log\,$g$ are primarily due to the distance error and amount to 0.06 dex. 
Lastly, microturbulent velocities are again based on the EW plot, where we restricted the measurements to the reliable regime of EW$<$180m\AA, 
both for the dwarf stars' lines and those measured in the more metal rich reference star. 
\subsection{Ionization equilibrium}
The average  [\ion{Fe}{1}/\ion{Fe}{2}], relative to the respective standard stars, of the entire RGB and TO sample is $-0.03\pm0.05$,which would indicate that ionization equilibrium in the CG stars is fulfilled. 
However, since once possible explanation for the departure from this equilibrium in the GCs studies in KM08 and KM10 were non-LTE (NLTE) effects, it is necessary to consider the difference of ionized and neutral species for the target  stars of either evolutionary status separately. 
\subsubsection{Red giants}
As was shown in KM08,  ionization equilibrium is not established in 
Arcturus, where neutral and ionized species differ by
$\varepsilon$(\ion{Fe}{1)}$-\varepsilon$(\ion{Fe}{2})=$-$0.08 dex. 
We remind the reader that the {\em differential} [\ion{Fe}{1}/\ion{Fe}{2}] abundance ratio in the 47 
Tuc giants is +0.08 dex, while the ionization imbalance in the more metal poor GC M 5 follows the opposite trend of an over-ionization by 
$-$0.12 dex (Fig.~2).  

In the present work we face the  similar situation as for the M 5 stars of KM10 in that the
iron ratio for the ionized species, differential to Arcturus, is higher than the neutral one by  $-0.11\pm0.02$ dex  on average. 
%
% 0.02 == 1$\sigma$/sqrt(3)
%

In the mean, ionization equilibrium could be enforced, at an [Fe/H] higher by 0.04 dex, by lowering log $g$ by 0.21 dex. 
As was elaborated in our previous works we start by resorting to systematic errors in mass, luminosity and/or T$_{\rm eff}$ 
to investigate the apparent over-ionization of iron in the GC stars relative to Arcturus.

If stellar mass was the source of the discrepancy in surface gravity, our targets would need to 
have a mass of 0.44 M$_{\sun}$ on average, which is an unlikely option. 
If changes in log $g$ were due to luminosity effects, the distance modulus of NGC~6397 would have to be fainter 
by 0.5 mag (thus 5$\sigma$ with our conservative error estimate from above), rendering it  more distant by at least 0.7 kpc. 
Given the good agreement of all the recent values in the literature this is also unrealistic. 
Likewise, the photometric accuracy of Kaluzny (1997) and the K-band magnitudes for the bright RGB stars are typically  better than 
0.02 mag.  
Moreover, the combined contributions of BC uncertainties (at $\sim$$0.05$ mag) and the published 
errors on the reddening of typically 0.01 mag amount to an overall luminosity-effect on log $g$ of
no more than 0.06 dex  which is clearly below the 0.21 dex change in gravity required to explain 
the observed lack of ionization equilibrium of iron. 

At a deviation of color- and excitation temperatures on our Arcturus scale of 91 K,  it seems 
adequate to invoke changes in T$_{\rm eff}$ as the source for the non-equilibrium. 
In fact, an increase in the temperatures of 43 K on average would we able to 
resolve the ionization imbalance in our stars (see also Table~6).  This increase in
T$_{\rm eff}$ is possible if we favor our photometric temperatures over the excitation
values, which seems possible if the correction to the color-temperature relations are
larger for [Fe/H] near $-$2.  Also, while the S/N of our spectra is easily sufficient
to well measure EWs in excess of 6.0 m\AA, it is conceivable that systematic measurement
errors for lines in the 6--15 m\AA\ range could occur, and thus affect the derived
excitation temperatures.
A 43 K discrepancy could be also accounted for by an increase in the reddening 
by 0.02 mag, which is, however, 2$\sigma$ of the current best estimates in the literature. 
These temperature changes would result in [Fe/H] 
abundances higher by 0.07$\pm$0.01 dex on average, which are in contrast to the low
random and systematic uncertainties derived here (Sect.~4.1). 
These arguments indicate that temperature effects cannot be ruled out as a possible contributor
to the non-equilibrium. 

As another possible explanation we explored the sensitivity of the [\ion{Fe}{1}/\ion{Fe}{2}] ratio 
to the adopted $\alpha$-enhancement in the atmospheres. Switching from the opacity distributions 
with an [$\alpha$/Fe] abundance ratio of +0.4 (AODFNEW) to the scaled solar composition (ODFNEW)
is well able to re-establish the ionization equilibrium in the cluster stars on average (see also Table~6) 
and we do in fact find that [\ion{Fe}{1}/\ion{Fe}{2}]$_{\rm ODF}=0.00\pm0.02$ dex. 
However, as we demonstrate in Section~4.2 (and Table~6), the average enhancement of the red giants of NGC~6397 
in the  $\alpha$-elements (O, Mg, Si, Ca, Ti) amounts to 0.39$\pm$0.03 dex, and 0.44$\pm$0.06 dex, if also  
the light elements Na and Al are included (see also Castilho et al. 2000; Lind et al. 2010). Therefore, lowering the atmospheric opacities to a Solar level is 
not a viable solution for the observed imbalance. 
%
% Mean O, Mg, Si, Ca, Ti (Ti I only) for all RGBs: 0.39 +- 0.03 (std/sqrt(N_elements * N_stars)) and 0.43+-0.06 for all plus Na, Al

Deficient Fe~I, relative to Fe~II abundances, is the oft-cited sign of NLTE over-ionization of
Fe~I atoms.  We note, however, that Korn et al. (2003) in a NLTE study of cool giants, and 
Korn et al. (2007) in an abundance study of NGC~6397 stars, found NLTE over-ionization corrections 
to Fe~I abundances of a mere 0.03 dex; however, this result depends on the adopted efficiency of
collisions with hydrogen.  On the observational side we note that differential line by line
abundances from Fulbright et al. (2007) show no trend in $\Delta$(FeI$-$FeII), at less than 0.05 dex,
over nearly 2.0 dex in [Fe/H] and 1000 K in temperature for cool RGB stars; similarly, Ti showed
a similar, small, $\Delta$(TiI$-$TiII) with no change over the same range of parameters.  It was
concluded that any NLTE effect is small in cool RGB stars; because these stars have quite low
temperatures there is an absence of UV flux in the atmosphere for over-ionization of metals.
We conclude that the difference in abundance between Fe~I and Fe~II lines in this work is unlikely
to have been caused by NLTE effects.

It is in principle conceivable that the lack of ionization equilibrium can be explained by an
anomalous He content of the NGC~6397 giants in that a helium-enrichment yields a simple shift in
the surface gravities. Following the formalism of Str\"omgren et al. (1982; see also Lind et al. 2011),
we can show that the necessary gravity change of 0.21 dex would require Y=0.65  in the atmospheres, 
which is about three times the  standard He admixture. 
In fact it appears likely that  the combined effects discussed above conspire to lead to the observed order of magnitude for the lack of differential 
ionization equilibrium in the giant stars. 

Finally, we note that Ti ionization equilibrium does not hold for the NGC~6397 giants, neither, with a mean difference $\Delta\varepsilon$(\ion{Ti}{1}$-$\ion{Ti}{2})=$-0.16\pm0.06$ dex.
The sense of this departure is the same as for iron, i.e., the ionized species yields higher abundances 
on average than the neutral species. 
\subsubsection{Turnoff stars}
In the TO stars we see the opposite trend, with a higher differential abundance of the neutral species, i.e.,  [\ion{Fe}{1}/\ion{Fe}{2}] = +0.09$\pm$0.08 dex\footnote{Note that the formal error on the mean is strictly zero so that we quote here the mean measurement error of 0.08 dex.}. Thus ionization equilibrium 
is marginally fulfilled in the hotter dwarf stars. 
Similarly, for titanium we find $\Delta\varepsilon$(\ion{Ti}{1}$-$\ion{Ti}{2})=$+0.05\pm0.03$ dex. 
As was found in KM08, the equilibirum already holds in the reference star itself, where 
$\varepsilon$(\ion{Fe}{1)}$-\varepsilon$(\ion{Fe}{2})=$-$0.02 dex.

In analogy to the above, this imbalance could be alleviated by an increase in surface gravity by 0.26 dex, corresponding to T$_{\rm eff}$ lower by 180 K. 
In this case, the new balance would be settles at a mean [Fe/H] of $-2.14$ dex -- consistent with the values found from the RGB sample. As we outlined in the previous section, neither changes in mass, distance, nor the He content could reasonably account for the observed gravity changes needed to resolve the ionization imbalance.   
Likewise, at a mean of $\alpha$-enhancement of the TO stars of 0.29$\pm$0.06 dex, the choice of the atmospheres' opacities is not a viable option. 
In conclusion, given the larger uncertainties of the TO stars' temperature scale (both compared to the more accurate T$_{\rm eff}$ of the RGB and the smaller errors on the TO's other  parameters) the choice of temperature is a feasible cause of the observed non-equilibrium in the two turn-off stars.
\subsubsection{Other  GCs}
In the following we combine the issue of departures of ionization equilibrium in the metal poor NGC~6397 stars with our findings in the more metal rich GCs 47~Tuc (KM08) and M5 (KM10). 
Since we found a systematic difference between the (one) TO and (eight) RGB stars in 47~Tuc as well as for the present data, we distinguish these stages in Fig.~2.
\begin{figure}[htb]
\begin{center}
\includegraphics[angle=0,width=1\hsize]{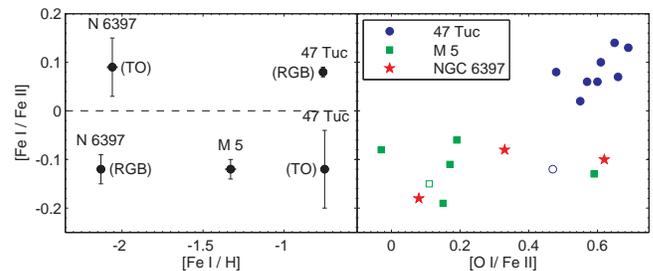}
\end{center}
\caption{Sense of the departure from ionization equilibria in the differential GC abundances
from KM08, KM10, and this work.  Note: the open circle indicates the TO star in 47~Tuc; the
open square indicates the AGB star, M5III-50, in M5.}
\end{figure}

While the sense of the non-equilibrium is inverted for the most metal rich CG in our sample, no linear decline of the departure is discernible in the 
{\em red giant stars};  moreover, the   [\ion{Fe}{1}/\ion{Fe}{2}]  ratio appears to level off towards the value of $\sim$0.1 dex below a certain (yet to be defined) 
metallicity threshold. 
Unfortunately, the M5 study of KM10 did not include any turn-off stars\footnote{The giant sample in M5, however, contains one AGB star with an 
 [\ion{Fe}{1}/\ion{Fe}{2}]  ratio  entirely consistent with the remainder of the RGB stars.} to bolster any systematics, but we note that either TO sample 
 in the metal rich 47~Tuc and the very metal poor NGC~6397 exhibits clearly opposite senses of the departure from equilibrium. 

An intriguing trend is then  seen in the right panel of Fig.~2: those stars with the largest enhancements in the O/Fe ratio also show the largest (positive) ionization imbalance  \ion{Fe}{1}/\ion{Fe}{2} -- the Pearson correlation coefficient between this ratio and the O-abundance ratio is 0.74$\pm$0.12. 
For 47~Tuc the deficiency of Fe~II, relative to Fe~I, cannot be explained by NLTE overioization;
possible explanations are outlined in KM08.  One scenario for explaining
the trend is that enhanced low-ionization metals, including $\alpha$-elements, in 47~Tuc provide
a source of electrons that significantly reduces the derived Fe~II abundance, akin to the effect
of increased gravity, while the Fe~I abundances are mostly unchaged.
\section{Differential abundance results}
Table~5 lists the abundance ratios as determined from the atmospheres with the parameters 
discussed above. All ratios are given relative to \ion{Fe}{1}, except for  [\ion{O}{1}], 
which,  due to a similar sensitivity to surface gravity, we state relative to \ion{Fe}{2}. 
\begin{deluxetable*}{rrrrrrrrrrrr}
\tabletypesize{\scriptsize}
\tablecaption{Abundance results}
\tablewidth{0pt}
\tablehead {\colhead{} & \multicolumn{3}{c}{\#13414} & & \multicolumn{3}{c}{\#8958} & & \multicolumn{3}{c}{\#7230} \\
\cline{2-4}\cline{6-8}\cline{10-12}
 \raisebox{1.5ex}[-1.5ex]{Ion} & \colhead{{[}X/Fe{]}} & \colhead{$\sigma$}& \colhead{N} & & \colhead{{[}X/Fe{]}} & \colhead{$\sigma$}& \colhead{N} & 
              & \colhead{{[}X/Fe{]}} & \colhead{$\sigma$}& \colhead{N}}
\startdata
{[}Fe\,I/H{]}               & $-$2.14 &  0.11 &  91 & & $-$2.17 &  0.12 &  80 & & $-$2.09 &    0.13 & 87 \\
{[}Fe\,II/H{]}              & $-$1.96 &  0.09 &   6 & & $-$2.09 &  0.09 &   5 & & $-$1.99 &    0.05 &  6 \\
{[}O\,I{]}\tablenotemark{a} &	 0.08 &  0.01 &   2 & &    0.33 &  0.12 &   3 & &    0.62 &    0.02 &  3 \\
Na\,I           	    &	 0.37 &  0.09 &   4 & &    0.28 &  0.22 &   3 & & $-$0.20 &    0.13 &  4 \\
Mg\,I           	    &	 0.47 &  0.17 &  5\ & &    0.51 &  0.23 &   5 & &    0.40 &    0.18 &  4 \\
Al\,I           	    &	 1.13 &  0.07 &   5 & &    1.04 &  0.13 &   6 & &    0.66 & \nodata &  1 \\
Si\,I           	    &	 0.43 &  0.15 &  11 & &    0.45 &  0.14 &   6 & &    0.45 &    0.17 &  8 \\
Ca\,I           	    &	 0.41 &  0.11 &  10 & &    0.38 &  0.13 &   9 & &    0.41 &    0.07 &  9 \\
Ti\,I           	    &	 0.29 &  0.15 &   8 & &    0.29 &  0.09 &   4 & &    0.39 &    0.18 &  5 \\
Ti\,II\tablenotemark{a}          	    &	 0.32 &  0.18 &   4 & &    0.37 &  0.15 &   4 & &    0.41 &    0.14 &  4 \\
\hline
\colhead{} & \multicolumn{3}{c}{\#13552} & & \multicolumn{3}{c}{\#365} & & & \\
\cline{2-4}\cline{6-8}
 \raisebox{1.5ex}[-1.5ex]{Ion} & {[}X/Fe{]} & $\sigma$ & N & & {[}X/Fe{]} & $\sigma$ & N  & & &  \\
\cline{1-8}
{[}Fe\,I/H{]}    & $-$2.03 & 0.14 & 53 & & $-$2.09 & 0.14 & 39 & & & \\
{[}Fe\,II/H{]}   & $-$2.12 & 0.18 &  7 & & $-$2.18 & 0.22 &  7 & & & \\
Mg\,I            &    0.11 & 0.11 &  3 & &    0.34 & 0.20 &  3 & & & \\
Ca\,I            &    0.42 & 0.20 &  7 & &    0.46 & 0.19 & 10 & & & \\
Ti\,I            &    0.15 & 0.40 &  4 & &    0.21 & 0.36 &  4 & & & \\
Ti\,II\tablenotemark{a}           &    0.15 & 0.22 & 21 & &    0.29 & 0.25 & 18 & & &  
\enddata
\tablenotetext{a}{Relative to \ion{Fe}{2}.}
\end{deluxetable*}
\subsection{Abundance errors}
In order to quantify the systematic errors on the chemical abundances due to 
uncertainties in the stellar atmosphere parameters, we performed the  
standard error analysis as in KM08 and KM10.  
Thus we varied  the parameters by the following conservative uncertainties 
and re-computed new abundances: (T$\pm$50K; log $g \pm$ 0.2 dex; $\xi \pm$ 
0.1 km\,s$^{-1}$; [$M$/H]$\pm$0.1 dex). The resulting changes in the  abundance ratios is displayed 
in Table~6, exemplary for one red giant and one TO star.  
\begin{deluxetable*}{l|rrrrrrrrrr}
\tabletypesize{\scriptsize}
\tablecaption{Error analysis for the red giant \#13414 and the turn-off star \#13552}
\tablewidth{0pt}
\tablehead{
\colhead{} & \colhead{} & \multicolumn{2}{c}{$\Delta$T$_{\rm eff}$} & \multicolumn{2}{c}{$\Delta\,\log\,g$} & \multicolumn{2}{c}{$\Delta\xi$} 
& \multicolumn{2}{c}{$\Delta$[M/H]} & \colhead{} \\
 &  \raisebox{1.5ex}[-1.5ex]{Ion}  & \colhead{$-$50\,K}  & \colhead{+50\,K} & \colhead{$-$0.2\,dex} & \colhead{+0.2\,dex} & \colhead{$-$0.1\,km\,s$^{-1}$} & \colhead{+0.1\,km\,s$^{-1}$} & \colhead{$-$0.1\,dex} & \colhead{+0.1\,dex} &  \raisebox{1.5ex}[-1.5ex]{ODF} 
 }
\startdata
& Fe\,I     				      &  $-$0.05 &    0.04 &	0.02 & $-$0.02 & $<$0.00 & $-$0.01 &	0.01 & $-$0.01 &    0.03 \\ 
& Fe\,II    				      &     0.05 & $-$0.07 & $-$0.08 &    0.05 & $<$0.00 & $-$0.02 & $-$0.02 &    0.01 & $-$0.08 \\
&{[}O\,I{]} 				      &  $-$0.01 &    0.01 & $-$0.06 &    0.07 &    0.01 & $<$0.01 & $-$0.02 &    0.03 & $-$0.07 \\
& Na\,I     				      &  $-$0.05 &    0.05 &	0.04 & $-$0.02 &    0.01 & $<$0.01 &	0.02 & $-$0.01 &    0.06 \\
& Mg\,I     				      &  $-$0.02 &    0.02 &	0.02 & $-$0.01 & $<$0.00 & $<$0.01 &	0.01 & $-$0.01 &    0.03 \\
 \raisebox{1.5ex}[-1.5ex]{\#13414} & Al\,I    &  $-$0.04 &    0.03 &    0.02 & $-$0.02 &    0.01 & $<$0.01 &    0.01 & $-$0.01 &    0.04 \\
& Si\,I     				      &     0.01 & $-$0.01 & $<$0.01 & $<$0.01 &    0.01 & $<$0.01 & $<$0.01 & $<$0.01 &    0.01 \\
& Ca\,I     				      &  $-$0.07 &    0.06 &	0.04 & $-$0.03 &    0.03 & $-$0.03 &	0.02 & $-$0.02 &    0.06 \\
& Ti\,I     				      &  $-$0.08 &    0.09 &	0.05 & $-$0.02 & $<$0.00 & $<$0.01 &	0.02 & $-$0.01 &    0.07 \\
& Ti\,II    				      &     0.01 & $-$0.02 & $-$0.05 &    0.03 &    0.03 & $-$0.03 & $-$0.01 &    0.01 & $-$0.04 \\
\hline
& Fe\,I    				      & $-$0.03 &  0.03 & $<$0.01 & $<$0.01 &    0.01 & $<$0.01 & $<$0.01 & $<$0.01 & $<$0.01 \\
& Fe\,II   				      & $<$0.01 &  0.01 & $-$0.06 &    0.08 &    0.01 & $<$0.01 & $<$0.01 &    0.01 & $<$0.01 \\
& Mg\,I    				      & $-$0.02 &  0.02 & $<$0.01 & $<$0.01 & $<$0.01 & $<$0.01 & $<$0.01 & $<$0.01 & $<$0.01 \\
 \raisebox{1.5ex}[-1.5ex]{\#13552} & Ca\,I    & $-$0.04 &  0.02 & $<$0.01 & $-$0.01 & $<$0.01 & $-$0.01 & $-$0.01 & $<$0.01 & $-$0.01 \\
& Ti\,I   				      & $-$0.05 &  0.04 & $<$0.01 & $-$0.01 & $<$0.01 & $-$0.01 & $-$0.01 & $<$0.01 & $-$0.01 \\
& Ti\,II  				      & $-$0.02 &  0.02 & $-$0.07 &    0.07 &    0.01 & $-$0.01 & $-$0.01 & $<$0.01 & $-$0.01 

\enddata
\end{deluxetable*}
We also computed atmospheres with Solar scaled opacity distributions, ODFNEW, 
which reduces the $\alpha$-enhancement in the input models by 0.4 dex. The effect of this  
variation is listed in the column labeled ``ODF''. 

In practice, the r.m.s. scatter of 30 K in the comparison of  T(V$-$K) versus excitation temperature 
indicates a 1$\sigma$ random error on either indicator of $\sim$21 K. 
If we  add in quadrature the systematic uncertainty of 30 K for Arcturus from KM08, we obtain a total error on the RGB T$_{\rm eff}$ scale of 37 K. 
For the TO stars we adopt a higher representative error of $\pm$150 K, consistent with the r.m.s. scatter of the different color-temperatures and the allowed range of H$\alpha$ fits.

By accounting for errors on  distance modulus, reddening, V-magnitude, 
BC and stellar mass (Sect.~3.2) we estimate a surface gravity uncertainty of 0.06 dex. 
Moreover, we assume a $\Delta\xi$ of 0.05 km\,s$^{-1}$, which is the allowed range to yield flat slopes 
in the EW vs. $\varepsilon$(\ion{Fe}{1}) plot, and a 0.05 error on the models'  metallicity [$M$/H]. 
Finally, we adopted an error on 0.1 dex on [$\alpha$/Fe], which  
corresponds to 1/4 of the difference when ODFNEW is used as opposed to the $\alpha$-enhanced atmospheres, 
and which is the typical  random scatter in the $\varepsilon(\alpha)$ 
abundances. 
Interpolating from Table~6, all these contributions are added in quadrature to obtain 
an {\em upper limit} for the final, uncorrelated systematic 
abundance uncertainty. Note, however, that the actual errors on our abundance ratios 
are probably smaller, due to the covariances of all atmospheric parameters (McWilliam et al. 1995).  
As a result, the total 1$\sigma$ systematic errors on the [\ion{Fe}{1}/H] and [\ion{Fe}{2}/H] 
abundance ratios of the RGB stars are 0.04 and 0.05 dex, respectively, and typically 0.02--0.06 dex for the 
light and $\alpha$-elements. 
Likewise, for the fainter TO stars, these ratios have systematic uncertainties of 0.11 (\ion{Fe}{1}), 0.16 (\ion{Fe}{2}), and 0.08 dex (other elements).  

Table~5 additionally lists the number of features $N$  that were measured to derive the elemental 
abundances, and the statistical error in terms of the 1$\sigma$ scatter of our measurements from 
individual lines. 
This error component is small compared to those from the atmospheric uncertainties 
for \ion{Fe}{1}, where a sufficient number of lines is detectable. For those elements with only a handful of lines present this 
random scatter will dominate the error budget. We find typical  mean random errors of 0.05 dex per line.
% == mean( scatter / sqrt(N) )
%
%
%
\subsection{Abundance ratios}
\subsubsection{Iron}
From our five targets   we derive a mean value of [\ion{Fe}{1}/H$]_{\rm LTE}$=
$-2.10\pm0.02\pm0.07$ dex (random and systematic error). The former, 
statistical error is simply the standard deviation of the mean. 
At 0.05 dex, the 
 star-to-star 1$\sigma$ scatter within NGC~6397 is small and consistent with the canonical notion of very homogeneous metallicities, at least 
as manifested in their iron content, of GCs (e.g., Carretta et al. 2009c). 
The larger systematic uncertainty above is mainly due to the contribution of the more uncertain temperature scale of the TO stars. 
The quoted mean is in excellent agreement with those found by Korn et al. (2007; NLTE) and Lind et al. (2009, 2011), averaged over all stellar types. 
Lind et al. (2008) find lower values around $-2.3$ dex, although on the lower RGB and with a broad scatter dependent on the evolutionary status. 
Our new abundance scale and the literature values cited above are more metal poor than those given by Castilho et al. (2000) and Carretta et al. (2009c) by 0.1 dex. 
As a main reason for any discrepancy we consider the differential nature of our study versus the primary use of $gf$-values in all other literature studies to date. 

At the respective, separate mean [Fe/H] of $-$2.13$\pm$0.02 (RGB) and $-$2.06$\pm$0.03 (TO), both evolutionary stages show iron abundances
different at the $2\sigma$ level (accounting for the random errors only). 
While Castilho et al. (2000) find no evidence of variations of the metal abundance with evolutionary status, Korn et al. (2007) and Lind et al. (2008) discuss 
their observed trends of [Fe/H] with T$_{\rm eff}$ in terms of atomic diffusion and mixing in the GC stars. 
On average, the Fe-abundance in their sample drops by $\la$0.2 dex across 800 K, albeit in the opposite trend as observed from our sample. The coolest stars 
of sample, however, reach only as low as $\sim$5000 K, which is still 800 K warmer than the RGB stars analysed by us. Thus the difference of 0.07 dex between
RGB and TO over 2000 K in the present work cannot be regarded as significant evidence for or against the acting of mixing processes in the sample GC stars. 

All of the studies above either enforced ionization equilibrium  and/or restricted their discussions to \ion{Fe}{1} as the primary species, so that we will not pursue the comparison of [\ion{Fe}{2}/H] with the literature any further. We note, however, that the \ion{Fe}{2} based abundance scale of Kraft \& Ivans (2003) favors a higher mean metallicity of this GC of $-2.02$ dex. 
\subsubsection{Alpha-elements --- Mg, Si, Ca, Ti}
Fig.~3 illustrates the derived abundance ratios in a statistical box plot, which shows the mean and interquartile ranges for each element. 
Much information about the detailed abundance distribution of NGC~6397 is found in the literature and we do not opt to 
repeat the main arguments from those sources (see e.g., Castilho et al. 2000; Korn et al. 2007; Lind et al. 2011), which  
classify NGC~6397 as a typical metal poor Galactic halo cluster with star-to-star scatter in the light elements (Lind et al. 2009). 
Also our differential results classify NGC~6397 as representative of the population  in that it is enhanced in the $\alpha$ elements Mg, Si, and Ca to the plateau halo 
value of +0.4 dex (note that Ti is less enhanced by ca. 0.1 dex). Oxygen shows a broad spread, which we will discuss in Sect.~4.2.3.
\begin{figure}[ht]
\begin{center}
\includegraphics[angle=0,width=1\hsize]{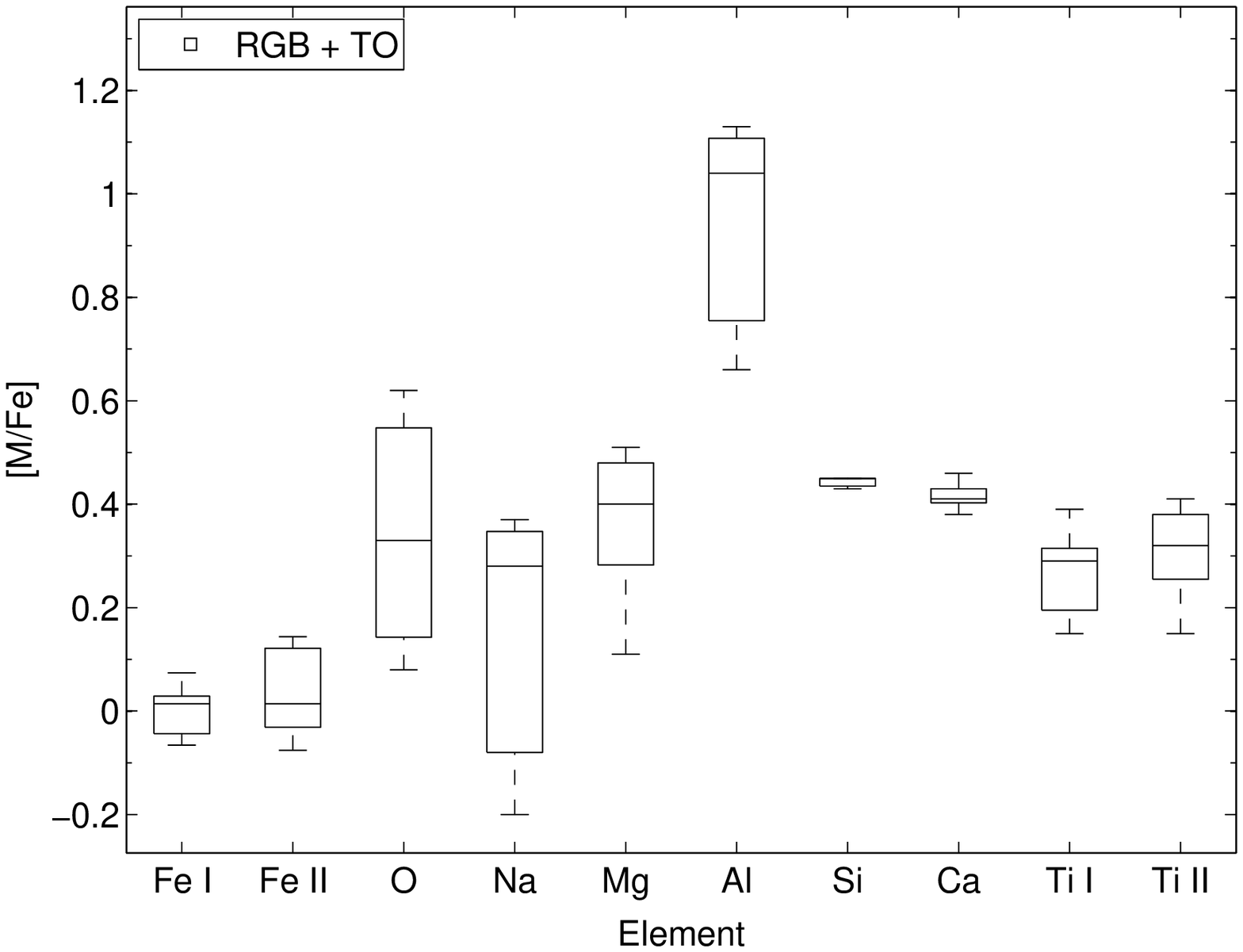}
\includegraphics[angle=0,width=1\hsize]{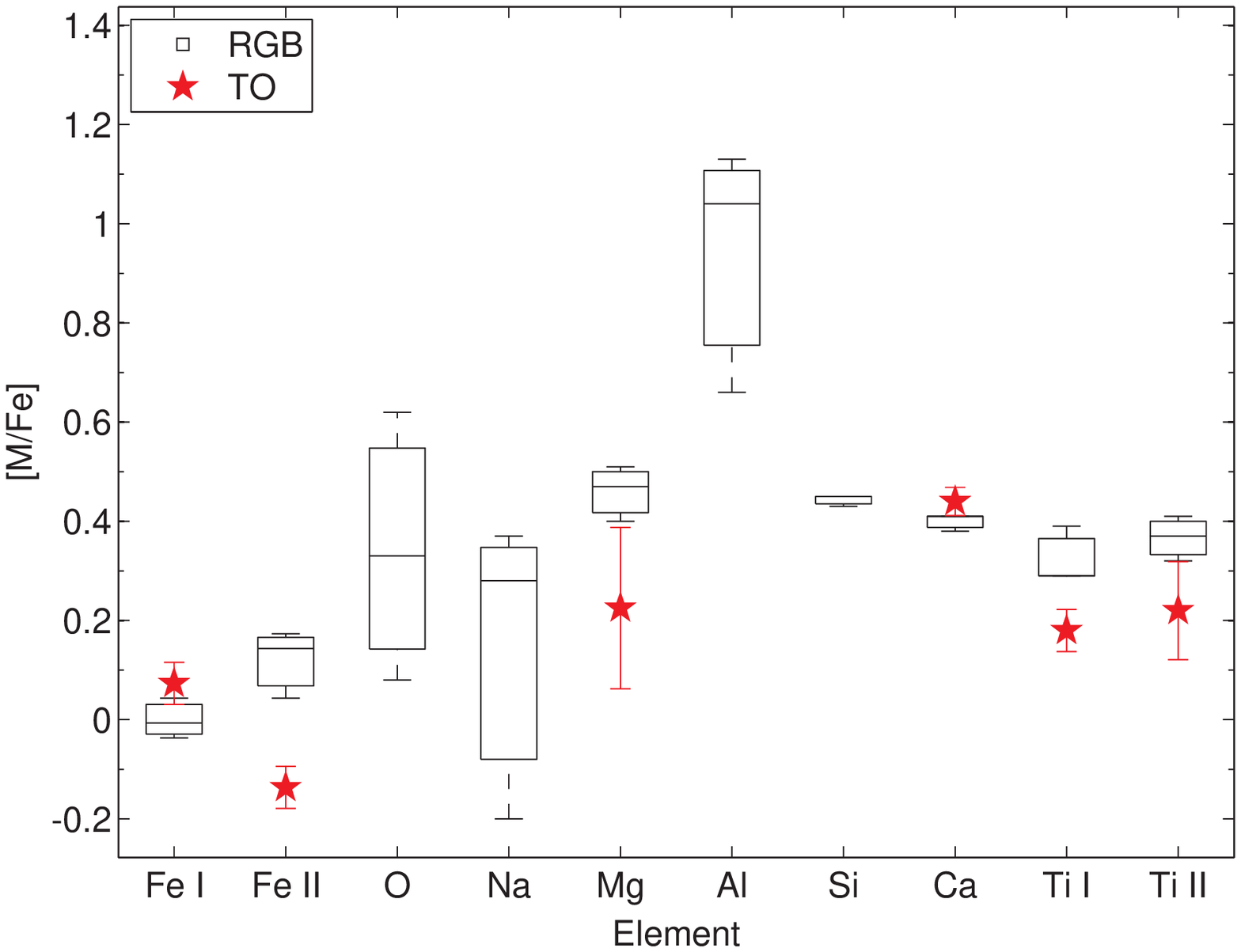}
\end{center}
\caption{Boxplots of the differential abundance results in our NGC~6397 sample. The top panel combines both RGB and TO sample, while the bottom panel shows the TO stars in relation to the RGB abundance ratios separately.}
\end{figure}

All other $\alpha$/Fe ratios in the RGB stars show a remarkable degree of homogeneity. Likewise, these elements do not show significant scatter between both TO stars, with the exception of Mg (and possibly \ion{Ti}{2}), which we shall address in the next section.  
As already addressed above, Korn et al. (2007) and Lind et al. (2008) report on the variation of NGC~6397's abundance ratios with evolutionary status, read: effective temperature. Thus we show in Fig.~4 the run of our measured abundances (relative to the respective mean values) with the atmospheric temperature. 
\begin{figure}
\begin{center}
\includegraphics[angle=0,width=1\hsize]{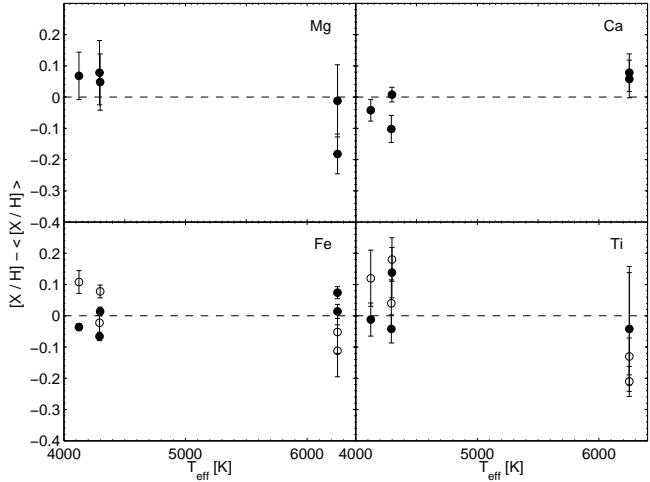}
\end{center}
\caption{Abundance ratios, relative to the respective mean, as a function of stellar effective temperature. For Fe and Ti open symbols refer to the ionized species.}
\end{figure}

The general consensus of the aforementioned studies was that a T$_{\rm eff}$ increase of $\sim$800 K incurs a decrease in the  [X/H] ratios for Mg and Ca by 0.07 and 0.17 dex, respectively, while the opposite trend is seen for Ti, albeit only marginally (Table~2 in Lind et al. 2008). Over the temperature range covered by those authors, these trends were well described by stellar atomic diffusion models (e.g., Richard et al. 2005). 
As already shown for iron, our data, reaching far cooler stars, do not show the same decrease but exhibit a marginal constancy. Evidence for falling abundance ratios at the hotter end is  caused by a lower value of [Mg, Ti/Fe] in TO star \#13552 at most. 
 
The $\alpha$-elements are predominantly 
produced in supernovae (SNe) of type II, that is, from massive, therefore 
short-lived, stars. While Mg and O are  formed in the hydrostatic nuclear burning in the 
SNe II progenitors, Si, Ca, and Ti are synthesized during the explosive phase of the SNe II 
(e.g., Woosley \& Weaver 1995). It can thus be expected to see different trends of in the abundance ratios of different $\alpha$-elements against each 
other  (e.g., Fulbright et al. 2007; KM10). 
This is shown in Fig.~5, where we follow the respective element ratios as a function of metallicity. 
Those plots and the following Figures illustrate our  data in comparison
with the differential abundances in the Galactic bulge (Fulbright et al. 2007) and the two GCs studied in KM08, KM10. 
\begin{figure}[htb]
\begin{center}
\includegraphics[angle=0,width=1\hsize]{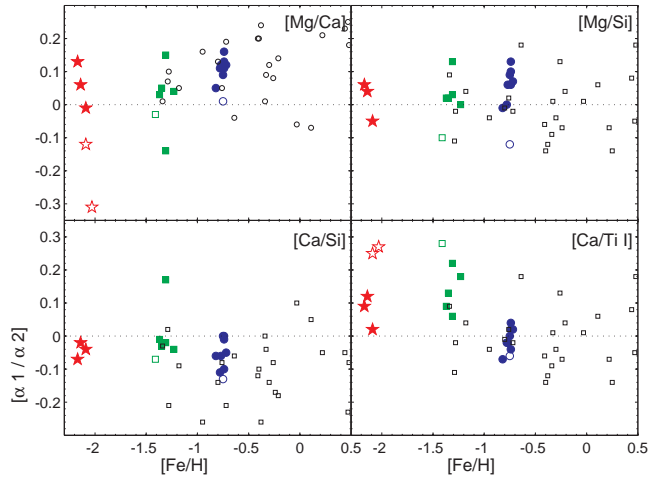}
\end{center}
\caption{Various $\alpha$-to-$\alpha$-element ratios for NGC~6397 (red symbols) and the GCs M5 (green squares) and 47~Tuc (blue circles) from our previous works. Red giants are designated by solid symbols, while TO stars (or, the AGB star in the case of M5) are indicated as open symbols. Also shown are the differential abundance ratios for Galactic bulge stars by Fulbright et al. (2007).}
\end{figure}

As a result, the [Ca/Si] ratios in all clusters are broadly consistent zero, as found in essentially all Galactic components, with only little scatter 
(r.m.s. of 0.03--0.08 dex). This confirms that the explosive $\alpha$-elements indeed 
trace each others' element trends. 
This holds for the observed [Mg/Si] ratios as well, which are compatible with zero in our differential analyses (at a mean of 0.02$\pm$0.03 dex). 

At a mean of $\sim$0.15 dex  (1$\sigma$-scatter of 0.05 dex), the [Ca/Ti] ratios in the metal poor GCs NGC~6397 and M5 are slightly higher than the values for field stars. It is interesting to note that both these [Ca/Ti] values are higher by about 0.15 dex than in the metal rich 47~Tuc  (KM08). 
As the [Mg/Ca] plot in the top panel of Fig.~5 shows,  this ratio is, for NCG~6397 and M5,  in full agreement within their scatter with the value of $\sim$0 found in the Galactic halo, bulge and disks. In fact, the 1$\sigma$ scatter in this element combination is larger than, e.g., in the Ca/Si and Ca/Ti ratio. 
It is, however, evident, that most of the scatter in the data from this work is due to different values as a function of evolutionary status. 

\subsubsection{Light elements --- O, Na, Al}
None of these elements could be measured in the TO stars, since the usual reliable transitions were too weak to be identified. On the other hand,  the strong Na D lines were still too strong for meaningful measurements and also the blue Al 3940, 3960\AA~Al lines were too strong and close to the broad Ca K feature in the reference spectrum of Hip~66815. Note that we did not apply any NLTE corrections to our Na and Al results.

The oxygen abundances in the red giants are based on the [\ion{O}{1}] 6300, 6363\AA~lines that we carefully 
deblended from telluric absorption, and the much weaker feature at 5577\AA. 

In previous works we had chosen to reject the stronger lines at 5682\AA, which are heavily blended 
with other metal lines in more metal rich stars. In the metal poor NGC~6397, however, such blends are negligible and we use a mean  from the 5682 and 6154\AA~lines as 
our final sodium abundance (Ivans et al. 2001). 

Both Na and O  show the broad range in their abundance ratios, in accord with the now well established light-element variations, in particular the Na-O anti correlation, in all GCs studied to date (Gratton et al. 2004; Carretta et al. 2009a,b), as also seen in NGC~6397 (e.g., Lind et al. 2011). 
Two of the red giants show [O/Fe] and [Na/Fe] ratios in the range of $\sim$0.1--0.4 dex, which are compatible with the majority of the GC stars
and consistent with the dominant (intermediate, ``I'') second generation of stars, enriched by the ejecta of (massive AGB or fast rotating) stars that could sustain proton-capture reactions in their interiors. 
One star (\#7230), on the other hand,  has a remarkably high oxygen-to-iron ratio and is accordingly depleted in sodium. This is attributable to a pure enhancement of an early generation of SNe~II  so that this star can be assigned to the ``primordial'' (P) cluster generation (Carretta 2009a).  
Although our sample size of three stars does not allow us to statistically investigate the fractions of stars in either population and thus to trace in detail the enrichment histories of the primordial and second stellar generations, we note that the ratio of P/I of 1:2 is fully consistent with the empirical ratio found in the majority of GCs irrespective of their metallicity (Carretta et al. 2009a).

Aluminum abundance ratios in the giants were measured from the moderately weak ($\la$32 m\AA) lines around 6696 and 7835\AA~and, for \#8958, from the weak 5557\AA~transition. 
As a result, we see evidence for a Mg-Al correlation and, in the top panel of Fig.~6, we parallel the Mg/Al correlation with the NaÐO anti-correlation discussed above. 
\begin{figure}[htb]
\begin{center}
\includegraphics[angle=0,width=1\hsize]{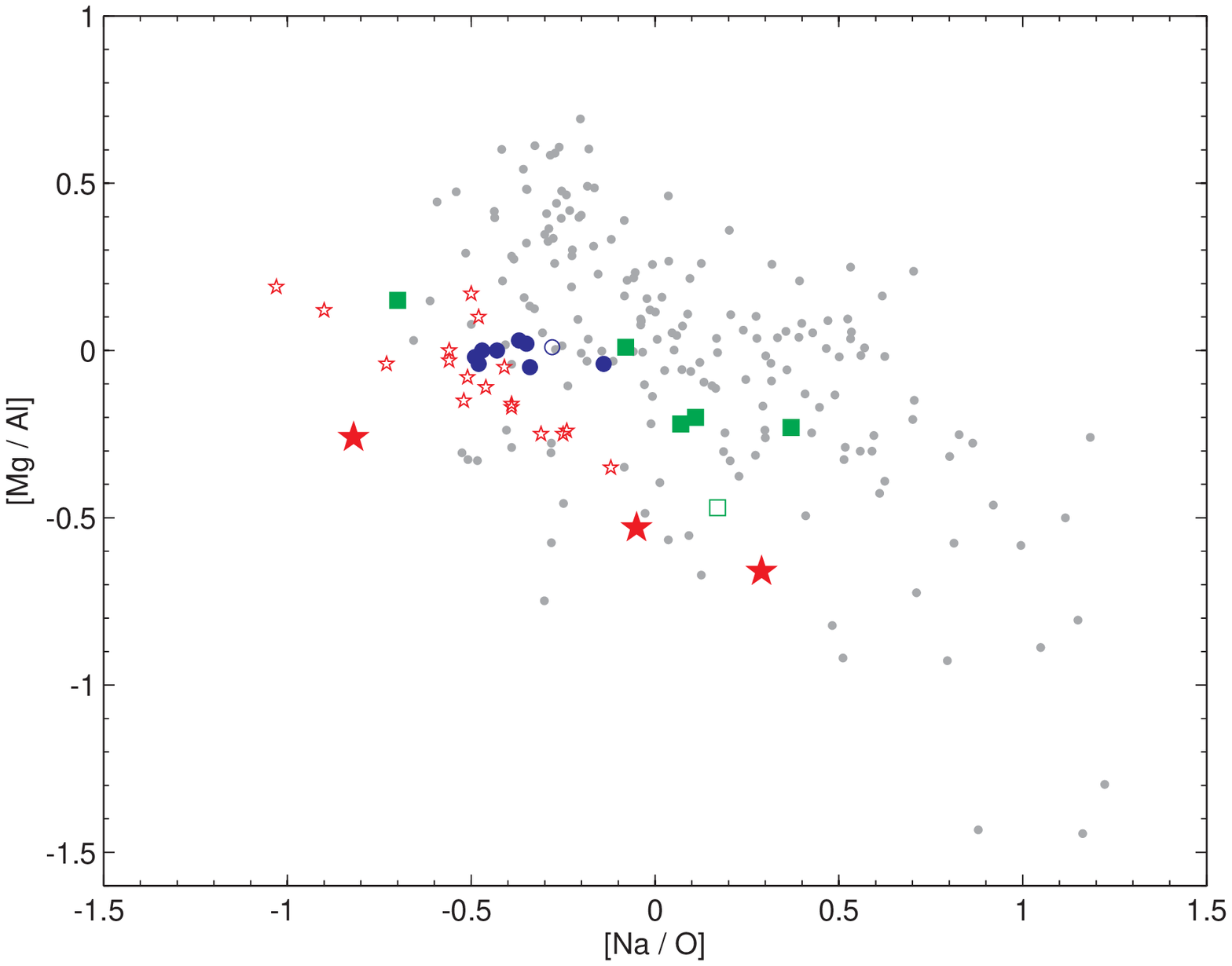}
\includegraphics[angle=0,width=1\hsize]{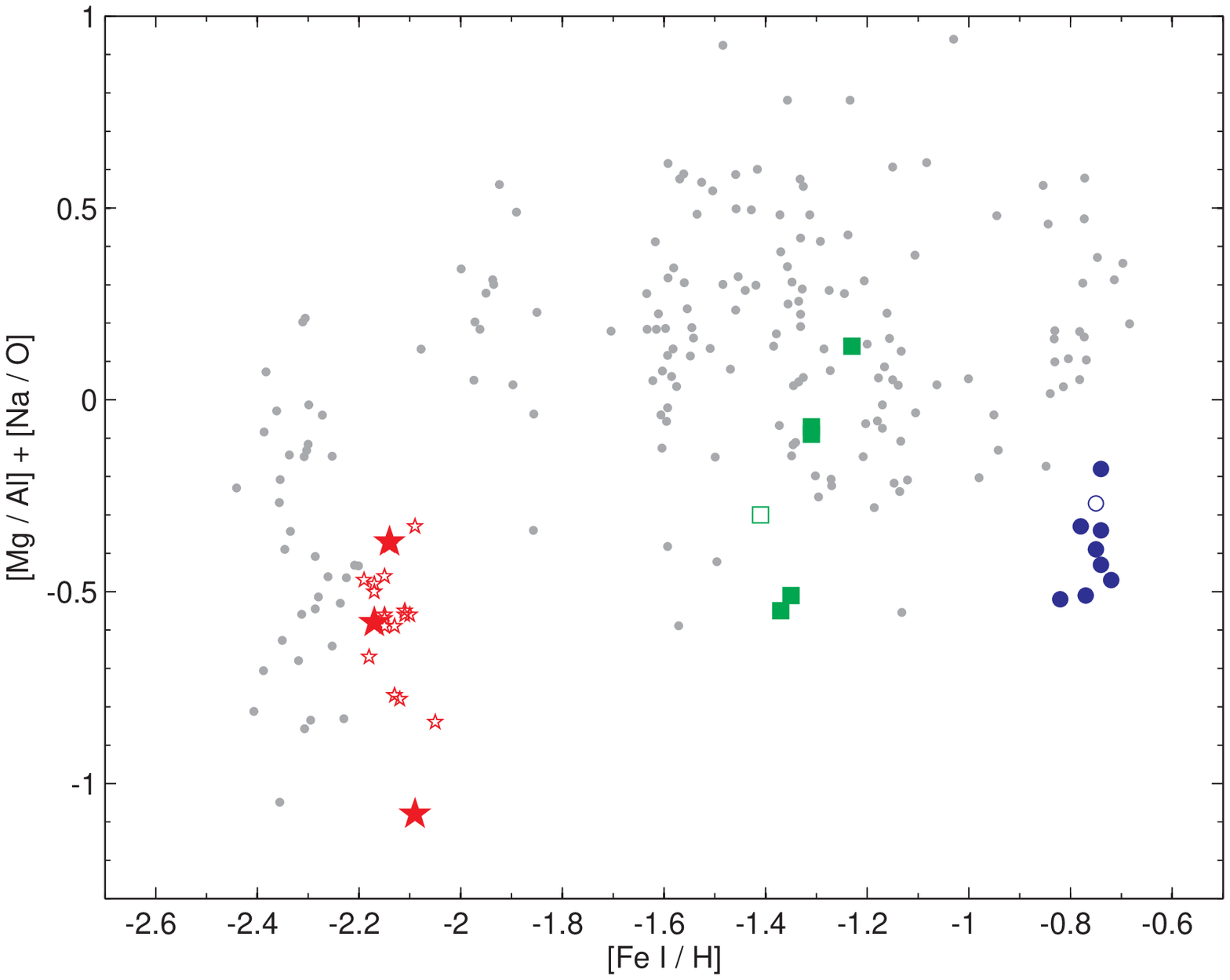}
\end{center}
\caption{The Mg-Al and Na/O correlation in red giants of NGC~6397 (red solid star symbols), M5, and 47~Tuc (same symbols as in Fig.~2). 
Data for 19 GCs are taken from Carretta et al. (2009a,b; open cyan circles). We also show as red open stars the measurements in red giants in  NGC~6397 by
 Lind et al. (2011).}
\end{figure}

As before, it is evident that \#7230, with the highest Mg/Al ratio, reflects the original, chemically unprocessed composition of this GC. 
Likewise, the different Mg/Fe ratio in the two TO stars may be indicative of their belonging to the different populations. A qualitative comparison of the Na D and Al~3940, 3960 line strengths indicates that, at the stronger Na features and slightly higher [Mg/Fe] of \#365, this object is also coined by the first SNe~II ejecta in the early enrichment phases of the GC (recall the very similar atmospheric parameters of these stars).
Overall, our differential abundances thus confirm the occurrence of star-to-star variations in the light elements due to the canonical proton-capture synthesis processes seen in other metal poor Galactic halo GCs.
\subsubsection{NLTE effects}
Fig.~6 (top panel) reveals an additional striking trend: at any given value of Na/O the stars in NGC~6397 show systematically lower Mg/Al ratios than the bulk of GCs studied in the literature. Also the data of Lind et al. (2011) are in agreement with this notion. The comprehensive data set of Carretta et al. (2009a,b) alone shows an indication of a detached sequence of stars with relatively low Mg/Al ratios. As it turns out, these belong to the metal poor M10 ([Fe/H]=$-1.56$; 2 out of their 10 stars), 
M15 ($-2.34$; 1/13), M30 ($-2.36$; 4/10), and M68 ($-2.23$; 9/12). 
We explore in the bottom panel Fig.~6 the sum 
$\Lambda\equiv$ [Mg/Al] + [Na/O], which, in other 
words,  simply combines the ratio of hydrostatic $\alpha$-elements [Mg/O] with the ratio of the proton-capture elements [Na/Al]. 
%	Lambda for "L"ight elements ... gotta be creative =]
%
The plot shows a clear dichotomy in that the mean and 1$\sigma$ scatter for GC stars with [Fe/H]$>-2$ dex amount to $\Lambda=0.18\pm$0.28 dex, while 
the metal poor stars below $-2$ dex have $\Lambda= -0.36\pm0.32$ dex -- a difference significant at the 80\% confidence level. 

Since both the literature and our data in this plot are based on LTE, we discuss here qualitatively 
NLTE corrections as a possible explanation for the observed discrepancy  (see, e.g.,  Fulbright et al. 2007)

{\em Oxygen:} Since  our O abundances are based on the meta-stable lines at 6300, 6363 \AA, the quoted abundance ratios  will only have negligible NLTE corrections. 
  
{\em Magnesium:} Andrievsky et al. (2010) perform an extensive analysis of Mg NLTE corrections. Although their calculations only cover warmer stars ($\ga$4500 K) at lower metallicities than in NGC~6397, we can qualitatively extrapolate the order of magnitude applicable to the cool red giants in the GC studies of interest. 
As the NLTE correction $\Delta$Mg (in the sense NLTE$-$LTE) is 0.10 dex at (T$_{\rm eff}$, [Fe/H])=(4600 K, $-$3 dex), and considering that $\Delta$ tends to decrease with decreasing temperature and for lower-metallicity stars, we estimate the correction not to be in excess of $\sim$0.1 dex for the metal poor GC stars. 

{\em Sodium: } For the more metal rich reference star, Arcturus, Fulbright et al. (2007) assume $\Delta$=$-$0.17 dex based on the calculations for the 5682, 5688\AA-lines of Takeda et al. (2003) and a smaller contribution from the 6154, 6160\AA~transitions. Following Takeda's et al. (2003) results for stars at lower [Fe/H] as in NGC~6397, we expect Na NLTE values lower by $\sim -0.3$ dex.  

{\em Aluminum:} Andrievsky et al. (2008) studied the strong resonance lines at 3944, 3962\AA, while our RGB Al/Fe ratios are based on the red 6696, 6698\AA-lines. 
If most of the effect on the resonance lines arises from over-ionization, then the red lines would show a similar
result and we can estimate the likely NLTE-corrections from those resonance lines near the stars closest to our giants' parameters  
as $\sim$0.4 dex.

In summary, a combination of the above NLTE effects on $\Lambda$ of $\Delta_{\Lambda} = % (0.1 - 0.4) + (-0.3 - 0.0) = -0.6
-0.6$ dex is a viable option to account for the systematically lower Mg/Al abundance ratios for metal poor stars with respect to their Na/O ratios. 
\section{A note on mass loss}
Most of the luminous giants in the present data set and in KM08 and KM10 showed notable red- and blue-shifted emission in the wings of H$\alpha$ line. 
Therefore, we follow, e.g.,  M\'esz\'aros et al. (2008) in  measuring the mass outflow velocities from 
the  bisector of H$\alpha$. In Fig.~7 we correlate the respective measure of these stellar winds with the 
luminosities that are based on the isochrones discussed in Sect.~3.2.1. 

\begin{figure}[htb]
\begin{center}
\includegraphics[angle=0,width=1\hsize]{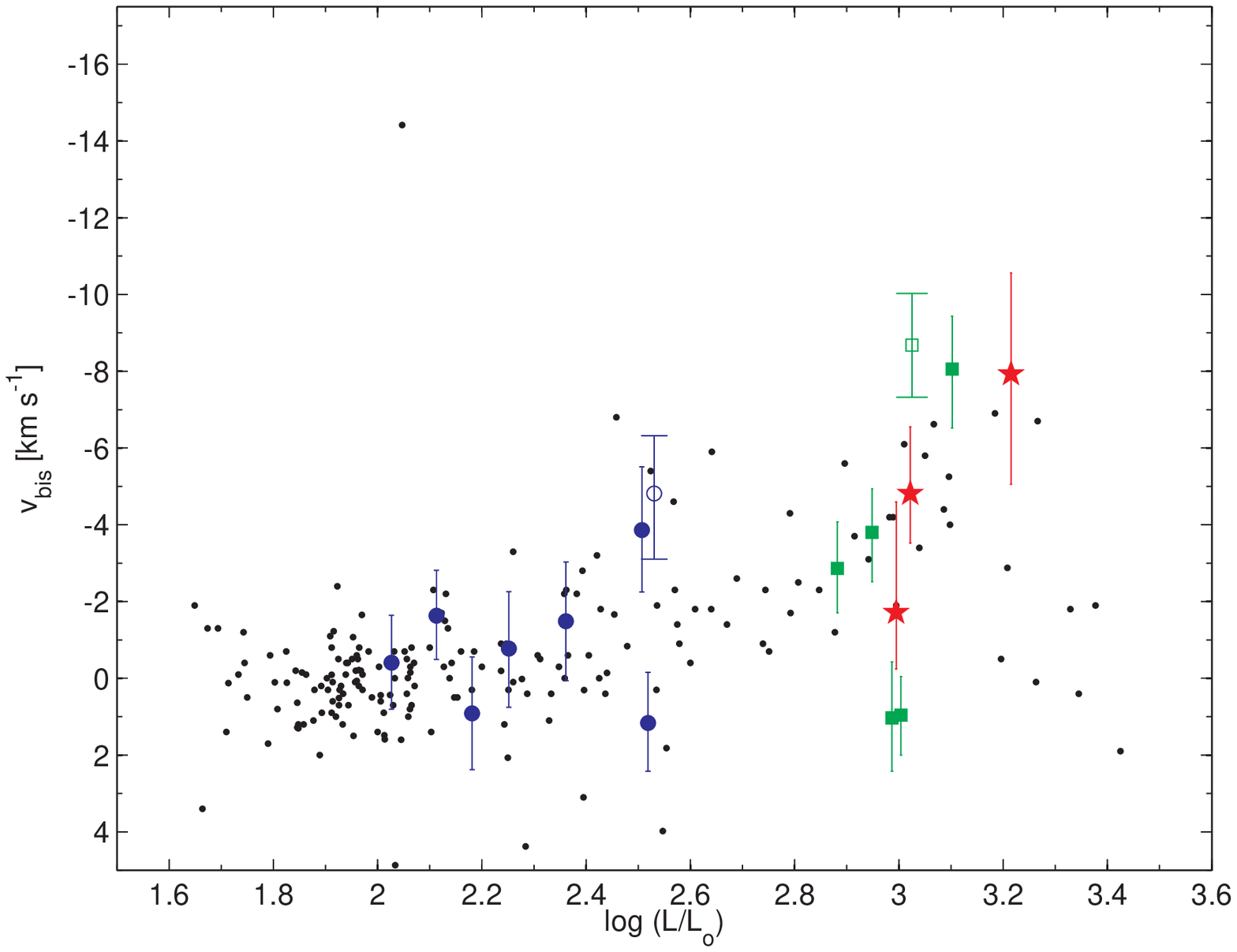}
\includegraphics[angle=0,width=1\hsize]{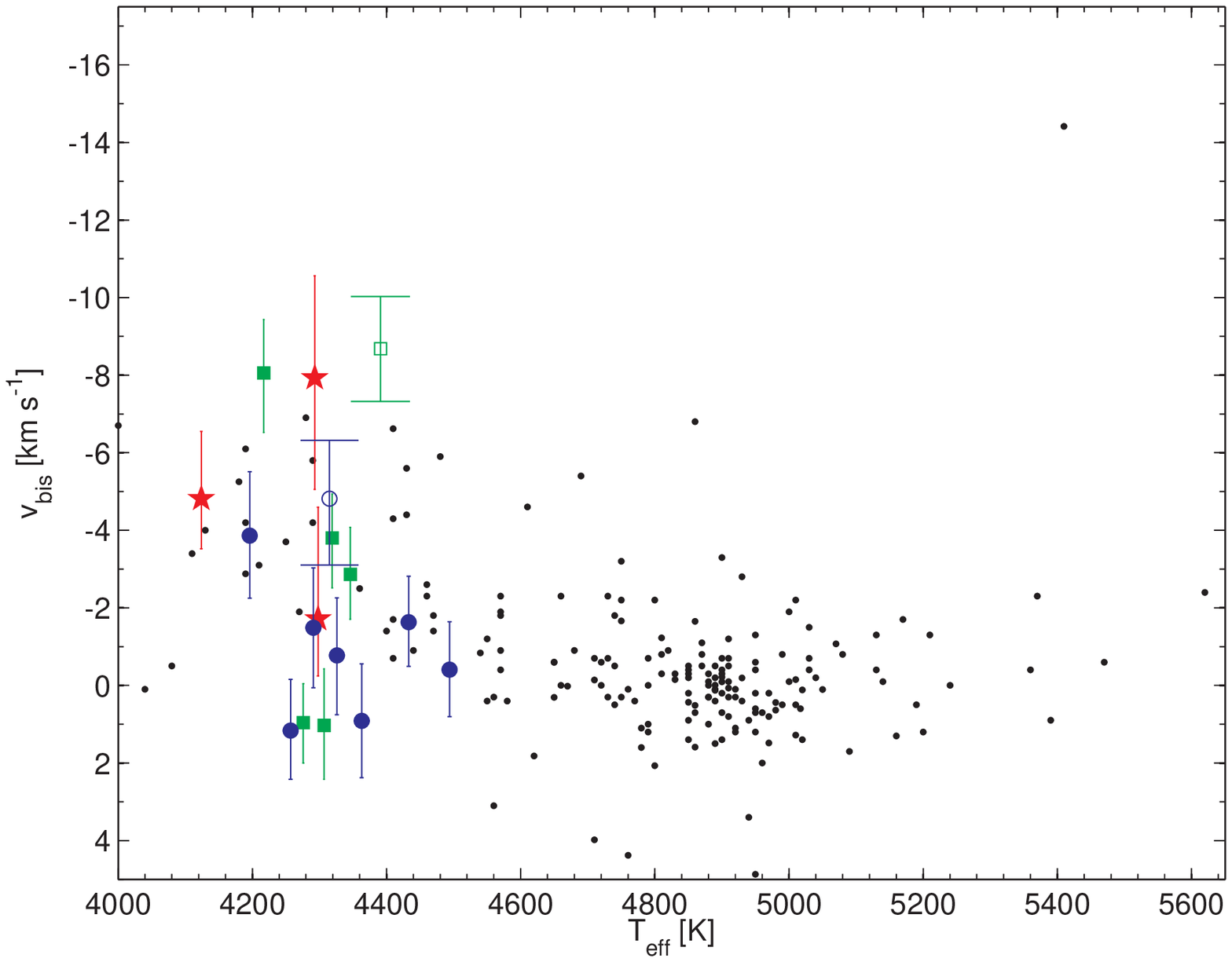}
\end{center}
\caption{Outflow velocities from the H$\alpha$ bisector. Shown are data from M\'esz\'aros et al. (2008, 2009) for three Galactic GCs and the analogous measurements in the spectra of 47~Tuc (KM08), M5 (KM10), and NGC~6397 (this work). Symbols are the same as in Fig.~2, where open symbols refer to the AGB stars.}
\end{figure}
The figure also shows results from three other GCs at [Fe/H] = ($-2.28$, $-$2.26, $-$1.54) from M\'esz\'aros et al. (2008, 2009).
Across the entire luminosity range covered in our samples the red giants in 47~Tuc, M5, and NGC~6397 are in good agreement with the literature data, thus confirming 
the empirical trends describing chromospheric expansions on the RGB. 
These are an onset of stronger winds at T$_{\rm eff}\la4500$ K, at luminosities above $\log L/L_{\odot} \ga2.5$, respectively. 
Moreover, the two AGB candidates in 47~Tuc (\#3) and M5 (M5III-50) clearly show the fastest outflow velocities amongst the respective cluster stars. 
This behavior is dependent on the stellar metallicity on the AGB -- the moderately metal poor M5 AGB star has an outflow velocity larger by 
4$\pm$2 km\,s$^{-1}$ than the metal rich equivalent in 47~Tuc. 
On the other hand, there is no systematic trend of the atmospheric velocities with cluster metallicity discernible in the RGB sample, as is substantiated by the 
comparison with the (moderately) metal poor GCs of M\'esz\'aros (2008, 2009).  
\section{Summary \& Discussion}
Here we have determined the chemical abundance ratios in three red giants and two TO stars in the metal poor 
Galactic halo globular cluster NGC~6397. 
Being a well-studied system, we indeed confirm that the majority of the $\alpha$-elements are enhanced to the +0.4 dex-halo plateau. Likewise, 
we find the canonical light element variations in O, Na, and Al indicating that one of our RGB targets and possibly one of the turn-off stars  were affected by an early 
stage of SNe II pollution. 

However, the present analysis vastly improves on the measurements of Fe, Na, Al, and $\alpha$-element-to-iron ratios 
(O, Mg, Si, Ca, Ti) by using line-by-line {\em differential} analyses relative to reference stars of similar stellar parameters (Arcturus and Hip~66815). 
This procedure serves to efficiently reduce uncertainties in atmosphere parameters and potentially erroneous atomic data. As a result, we derive an accurate 
mean LTE iron abundance of $-2.10\pm0.02\pm0.07$ dex, in agreement with recent high-resolution measurements based on $gf$-values. 
We note, however, that some studies have revealed star-to-star variations as a function of stellar evolutionary type, which do not significantly witness in our data. 
Unfortunately  a star-by-star comparison with other sources is prohibited by the lack of overlapping stars between these samples. 
However,  other literature data place this GC at an [Fe/H] higher by $\sim$0.1 dex. Already in our earlier comparison of the newly established  abundance scale 
of Koch \& McWilliam we noted such differences (e.g., 47~Tuc being more metal poor than on other scales; KM08), while M5 was found in excellent agreement 
with the literature (KM10). Thus we emphasize again that, there is no systematic offset that defines the differential abundance scale, but that any comparison of GC 
scales must be considered on a case-by-case basis.  
In particular, with the three GCs studied in this differential manner, this homogeneous abundance scale now covers a broad range of Galactic GCs' metallicities, from 
below $-$2 dex to the metal rich 47~Tuc, at $-$0.76 dex. 

\vspace{0.4cm}
\acknowledgments
We gratefully acknowledge funding for this work from a NASA-SIM key project grant, entitled ``Anchoring 
the Population II Distance Scale: Accurate Ages for Globular Clusters and Field Halo Stars''. 
AK thanks the Deutsche Forschungsgemeinschaft for funding from  Emmy-Noether grant  Ko 4161/1. 
This research has made use of the NASA/ IPAC Infrared Science Archive, which is operated by the Jet
Propulsion Laboratory, California Institute of Technology, under contract with the National Aeronautics
and Space Administration.

\end{document}